# Fiber neural networks for the intelligent optical fiber communication signal processing

Yubin Zang,[1] Zuxing Zhang,[1,*] Simin Li[2], Fangzheng Zhang[2], Hongwei Chen[3,4,*]

[1]*Advanced Photonic Technology Lab, School of Optoelectronic Engineering, Nanjing University of Posts and Telecommunications, Nanjing, China, 210023*
[2]*College of Electronic and Information Engineering and the Key Laboratory of Radar Imaging and Mircowave Photonics, Nanjing University of Aeronautics and Astronautics, Nanjing, China, 211106*
[3]*Department of Electronic Engineering, Tsinghua University, Beijing, China, 100084*
[4]*Beijing National Research Center for Information Science and Technology, Beijing, China, 100084*
*\*zxzhang@njupt.edu.cn;chenhw@tsinghua.edu.cn*

**Optical neural networks have long cast attention nowadays. Like other optical structured neural networks, fiber neural networks which utilize the mechanism of light transmission to compute can take great advantages in both computing efficiency and power cost. Though the potential ability of optical fiber was demonstrated via the establishing of fiber neural networks, it will be of great significance of combining both fiber transmission and computing functions so as to cater the needs of future beyond 5G intelligent communication signal processing. Thus, in this letter, the fiber neural networks and their related optical signal processing methods will be both developed. In this way, information derived from the transmitted signals can be directly processed in the optical domain rather than being converted to the electronic domain. As a result, both prominent gains in processing efficiency and power cost can be further obtained. The fidelity of the whole structure and related methods is demonstrated by the task of modulation format recognition which plays important role in fiber optical communications without losing the generality.**

**Introduction.** The surge of artificial intelligence (AI) has deeply and broadly changed the way of scientific researches including fields of photonics and optics. On one hand, various types of optical neural networks which designing photonic structures toward AI has been proposed to further increase computing speed and power efficiency. On the other, the neural network itself can be applied to establish fast and accuracy photonic or optic models in return.

As is listed in Fig.1, in fields of optical and photonic neural networks, several mile-of-stone structures like Mach-Zehnder Interferometer (MZI) structured photonic neural network [1], deep diffractive neural network [2] and etc have been put forward since 2017 [3-7]. In fields of AI-based optical or photonic models, rather more diverse cases can be found. Not only has AI permeates into the researches like optical membrane design, photonic waveguide optimizing, but also the fields of optical communications. Particularly, important AI-based fiber models including BiLSTM [8], GAN [9], multi-head attention[10, 11], Fourier operator models [12] and etc [13-16] have been proposed in recent years.

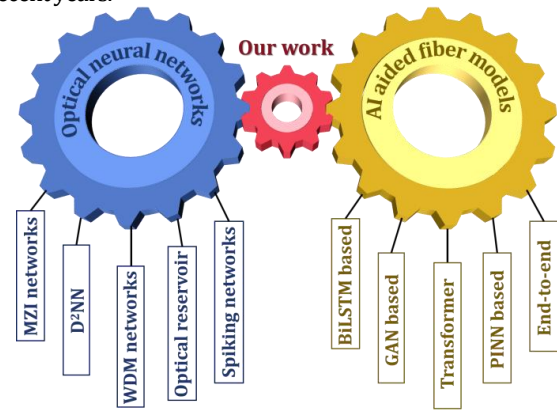

**Fig. 1.** Backgrounds and motivations of the work proposed.

Though both branches of AI optics as illustrated above have developed profoundly and rapidly, seldom has optics neural networks and AI-based optics models deeply integrated into each other. As in the disciplinary fields between AI and fiber communications, though optical neural networks and AI-aided fiber models have been both proposed at the early age, rarely has optical neural networks been applied into addressing optical fiber transmission problem. Therefore, in this letter, the fiber neural network and its related signal processing method towards addressing fiber communication signal processing problems is proposed. Once adopted, the fiber can not only realize signal transmission but also signal computation. It shred the hope that transmitted signals inside the core of fiber can be automatically processed all through optical domain instead of being converted into electronic domain and then processed. The fiber system will become the distributed computational transmission media from this point of view and will thus posses both high computing efficiency and low power consumption.

This letter will first introduce both the background and motivation of proposing the fiber neural network towards optical fiber communication signal processing. Without losing generality, modulate format recognition (MFR) which is signification is adopted to demonstrate the fidelity of the network. Under this circumstance, both the structure establishment and the dataset collection for MFR will be illustrated in the second part. Then, simulation setups and training procedures will be described in detail in the third part. All results and analysis will be developed in the fourth part. Conclusions and future work will be discussed in the last part.

**Network's structure and dataset.** The fiber optical neural network structure contains one input branch, one output branch and an optical computing ring. Components mainly include laser source, single mode fiber (SMF), erbium-doped fiber amplifier (EDFA), amplitude modulators (AM), dispersion compensation fiber (DCF), photodetector (PD) and etc all construct the whole structure as is shown in Fig.2. The function of input branch is to generate stretched light pulses to be modulated and the function of output branch is to convert the signals from optical domain into electronic domain for further processing and analyzing.

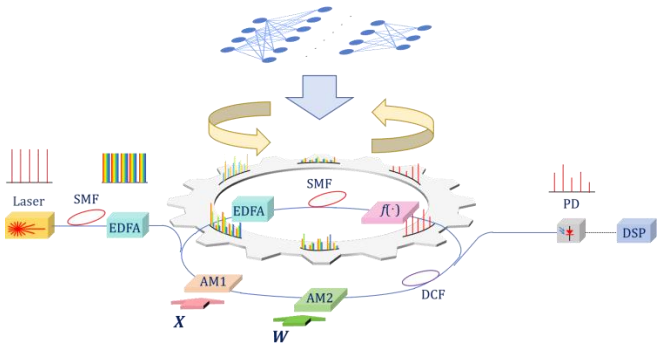

**Fig. 2.** Structure and principle of fiber optical neural network.

The computing ring in the model is to implement the configurations of the theoretic neural network model. The total mechanism lays down on the implementation of linear computations and nonlinear activation since most theoretic neural network models such as fully-connected networks, convolution networks or long-short term memory networks contain these two types of calculation.

When it comes to linear computation, it can firstly be viewed as matrix multiplication. Then, this multiplication can be further viewed as several multiplications between the row vector of the left-sided matrix with the column vector of the right-sided matrix. Under this circumstance, matrix multiplications can be realized by successively conducting vector multiplications in order. As for the computing ring, AMs, EDFA, SMF, DCF can implement this vector multiplication operation. When light travels into each loop of the ring, two vectors which are prepared to conduct the multiplication will firstly be modulated onto the peaks of light pulses, then DCF will take use the dispersion effect to compress the pulses so as to implement the vector multiplication. Later, matrix multiplications (linear computations) can be obtained by implementing each vector multiplication via loop in order.

Nonlinear activation can be realized via reconfigurable optical attenuation device which also exists in the optical computing ring structure. This device can realize different transmission ratio for different energy of light so as to implement the any types of nonlinear activation functions. Without losing the generality in this task, the activation function to be implemented is chosen to be 'Purelin' which has been broadly used in today's most theoretic neural networks.

The task for demonstrating this optical neural network's ability to process optical fiber communication signals is MFR since it has been one of the most frequently occurred problems in optical fiber communications. More specifically, MFR essentially requires the intelligent system to categorize different modulation formats according to the provided statistical information originated from the transmitted modulated signals. In this letter, three different modulation formats-OOK, PAM and PSK, each with four types of statistical attributes-algebraic mean, variance, variation and geometric mean are adopted to discern. Under this circumstance, the dataset of MFR consists of the four statistical attributes as the input and the classification of modulation formats which utilizes the 'one-hot' encoding as the label. In total 150 samples exist in the hole dataset and were separated into training dataset and testing dataset at the ratio 8:2.

In order to maintain the best activation states of each neuron, raw data need to conduct normalization before entering. As for this task, the normalization is determined to be Minmax normalization which maps the raw data into the ranges between -1 and 1. This normalization method has also been commonly used in other theoretic neural networks and has been one of the standard normalization types among all.

**Simulation setups and procedures.** For the establishment of optical fiber neural networks, in total three steps need to be done. First, the theoretic neural network model should be chosen. Second, hyper-parameters like loss function, learning rate, drop out ration need to be determined and then the network model needs to be appropriately trained to bridge the connection between the attributions input and the categorization labeled. At last, neurons' weights from the optimized theoretic neural network model should be transferred into modulated signals in the optical fiber neural network frame.

As for the chosen of theoretic neural network model, in order to alleviate the load of fiber optical neural networks, it is determined to be a three-layer fully-connected neural network, with 3, 6 and 4 neurons in each layer. The nonlinear function for the network is chosen to be 'Purelin' at the end of each layer.

When it comes to the training and optimizing the theoretic fully-connected neural network, hyper-parameters such as the loss function, optimization algorithm, learning rate and etc should be determined in advance. For this type of task, either normalized mean square error (NMSE) or cross entropy (CE) can be utilized as the loss function. For learning rate, it is chosen to be $10^{-3}$. The optimization algorithm is chosen to be adaptive momentum stochastic gradient descend method (ADAM) which has been frequently taken into use in the training of neural networks. In total 120 samples from the training dataset are injected into the fully-connected neuron network to let it establish the relations between statistical attributes with the correct modulate format category.

The final step for establishing the optical neural network is to map the neurons' weights from the optimized fully-connected neural network model to the optical fiber neural network frame. As for the theoretic fully-connected model, it can be concluded that both linear computations and nonlinear activation need to conduct between each layer. If the matrix $W$ and $b$ represents the neurons' weights and biases between each layer, $f(\cdot)$ represents the nonlinear activation function between each layer, $x$ and $y$ represent the network's inputs and outputs respectively, then the theoretic fully-connected neural networks can be described mathematically as

$$y = f(W_{23}f(W_{12}x+b_{12})+b_{23}), \quad (1)$$

where the scale of $W_{12}$, $W_{23}$, $b_{12}$, $b_{23}$ is 6×3, 4×6, 6×1, 4×1 respectively while the scale of $x$ and $y$ equals 3×1 and 4×1 which are in consist with the scale of the attributes input and the category output respectively.

For the processing convenience, if $b_{12}$ is positioned at the last column of the weight matrix $W_{12}$ and 1 is add at the bottom of the input vector $x$, then Eq.(1) can be simplified as

$$y = f(W_{23}f(W^e_{12}x^e)+b_{23}), \quad (2)$$

in which $W^e_{12}=[W_{12},b_{12}]$ and $x^e=[x,1]^T$. The same simplification can be conducted for the computations between the hidden layer and output layer. At last, Eq.(2) can be further simplified as

$$y = f(W^e_{23}f^e(W^e_{12}x^e)), \quad (3)$$

where $W^e_{23}=[W_{23},b_{23}]$ and $f^e(W^e_{12}x^e)=[f(W^e_{12}x^e),1]^T$.

Since the optical fiber neural network takes 'Purelin' as its nonlinear activation function, modified weight matrices $W^e_{12}$ and $W^e_{23}$ can be further combined as $W^e=W^e_{12}W^e_{23}$. Under this circumstance, Eq.(3) can be expressed as

$$y = W^e x^e, \quad (4)$$

The following and the most important work is to map both $W^e$ and $x^e$ into modulated signals in optical fiber neural network. Since optical fiber neural network utilizes intensity modulator which can not directly modulate negative weights onto the light while both $W^e$ and $x^e$ contain negative values due to the training process and normalization operations, extra actions need to be taken to help the fiber optical neural network deal with negative values. Here, the matrix symbol separation method that divides both $W^e$ and $x^e$ into positive components $W^{e+}$, $x^{e+}$ and negative components $W^{e-}$, $x^{e-}$ was adopted. In this case, Eq.(4) can be rewritten as

$$y=(W^+-W^-)(x^+-x^-)=(W^+x^++W^-x^-)-(W^+x^-+W^-x^+), \quad (5)$$

As can be observed from Eq.(5), Positive components keep positive values of the original matrices or vectors while taking 0 for negative values. Negative components take the opposite numbers of the negative values of the original matrices or vectors while taking 0 for positive values. When compared Eq.(5) with Eq.(4), it can be indicated that four matrix multiplication computations need to be done by the fiber optical neural network in order to implement the original one matrix multiplication in the theoretic full-connected neuron network.

Modulated signals mapping follows the transmission rules of the fiber optical frame. For each flattened-peaked light pulse, if $E_s(\cdot)$ and $E(\cdot)$ represent the waveform of electric field of the light pulse after and before the modulation and $s(\cdot)$ represents the modulated signal, then modulation operation can be mathematically expressed as [17, 18]

$$E_s = s(T)E = E\sum_{i=1}^{N}w_i x_i \times Rect_{\Omega\beta_2 l/N}\left\{T-(2i-1-N)\frac{\Omega\beta_2 l}{N}\right\} \quad (6)$$

where the index $i$ marks either the order of the element $w$ in weight matrix or $x$ in the input vector. $\Omega$ describes the angular bandwidth of the light pulse. $\beta_2$ and $l$ represents the second-order propagation constant which is related with chronic dispersion and the length of SMF. $T$ is the time-latent coordinate and $Rect(\cdot)$ represents the rectangular shaped function whose value equals one at the maxima and width is marked at its footnote. Under this circumstance, the results can be obtained at the end of each optical loop in the optical fiber neural network as

$$\mathcal{E} = \gamma\sum_{i=1}^{N}w_i x_i \quad (7)$$

where $\mathcal{E}$ represents the compressed pulse energy and $\gamma$ is the coefficient.

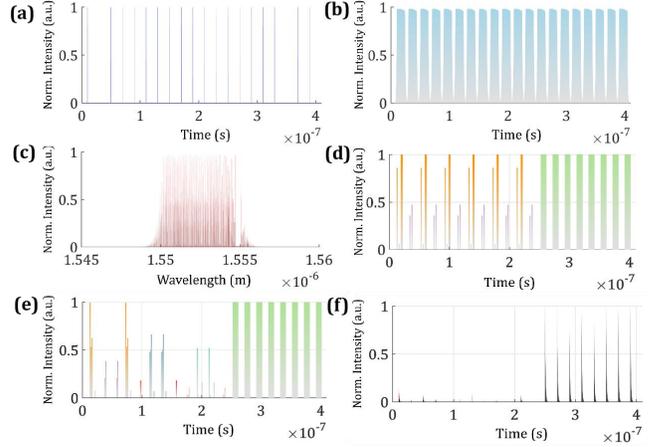

**Fig. 3.** Important signals in the fiber neural network. (a) After the laser; (b) laser spectrum; (c) before modulation; (d) $X$; (e) $W$; (f) received signals

---

**Results and analysis.** After the establishing of the fiber optical neural network, 30 samples, with 10 samples from each modulation format category that the neural network has never seen forms test dataset to test the real ability of discern.

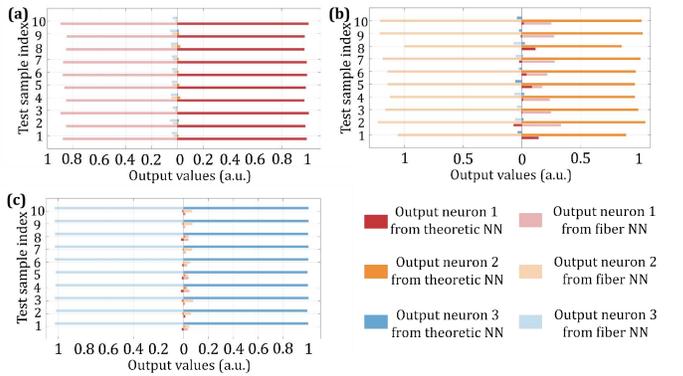

**Fig. 4.** Outputs from the theoretic neural network model and the fiber neural network. (a) OOK signals in test dataset; (b) PAM signals in test dataset; (c) PSK signals in test dataset.

---

In order to analyze the difference, important signals from each procedure of computation are depicted from Fig.3. Both time-domain signal waveforms, spectrum, waveforms before modulation, weights and inputs' signals compressed waveforms and results obtained can all be observed. As is shown in Fig3(a)-(c), the larger bandwidth, the easier for the dispersion to stretch the pulse. Here, the interval between each pulse is set to be 20ns while the time domain width of the pulse is 12ns. From Fig.3(d) and Fig.

3(e), organization of modulated signals with weights and inputs can be clearly observed. From the left to right, the pulses can be divided into three groups dealing with the multiplications between the first, second and third row vector of $W$ and $x$. For the first group, each pulse sub-sequentially modulated with the first row vector from $W^{e+}$ and the input vector $x^{e+}$, the first row vector from $W^{e-}$ and the input vector $x^{e-}$, the first row vector from $W^{e+}$ and the input vector $x^{e-}$, the first row vector from $W^{e-}$ and the input vector $x^{e+}$. The unmodulated pulses are references and will be measured to calculate the coefficient γ in Eq.(7) to help obatin the final results. Fig.3(f) shows the pulses after the propagation of DCF. It can be indicated that when stretched pulses with modulated signals travel through DCF, their separated frequency complements will overlap with each other due to the dispersion compensation effect. As a result, pulses will be compressed whose energy are closely related with the multiplication of the modulated signals as described in Eq.(7).

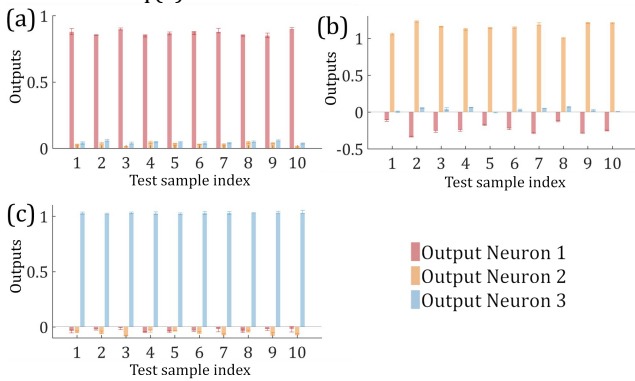

**Fig. 5.** Recognition outputs from the fiber neural network with noise. (a) OOK signals in test dataset; (b) PAM signals in test dataset; (c) PSK signals in test dataset

---

Differences in outputs from the theoretic model and fiber neural network can be found from Fig.4. They are mainly caused by the unflatness of stretched pulses, the pulse compression process and the mapping between pulses' energy into final results. The unflatness of stretched pulses which can be observed from Fig.3(b) causes the inaccurate of modulating. For pulse compression process, the ideal addition are hard to reach since there exists no perfect DCF which can completely compress the width of the stretch pulse into 0. In this case, the error may inevitably occur. The convert of pulses' energy into final results can also cause the error. It is because the compression results are closely related with pulses' shapes which are different since different pulses carry different weights or inputs. In conclusion, though errors exist, the results are acceptable as long as they does not deviate too far from the ideal ones.

Noise caused by EDFA and PD in the fiber optical neural network can have influence on final discern results as well. In order to test how noise affect the fiber neural network's performance, noise figure of EDFA is set to be 4dB, different noise like thermal noise, shot noise from PD are also added as well. Error bars are adopted to describe the results affected by the noise in Fig.5. It can be concluded that though errors exist, they does not exceed the boundary that can cause the mis-classification. This indicates that the fiber optical neural network can posses the relatively strong ability of resisting noise. Several possible reason for it can be obtained after carefully analysis. First, though signals are computed and propagated under the interference of noise from EDFA, the accumulation of noise can not affect much since pulses are compressed by DCF in each loop. The DCF performs as a low-pass filter which can cancel out portions of noise. Then, at the output branch of the fiber neural network, though noise from PD can not be neglected when turning optical signals into electronic signals, addition and subtraction between each pulses groups are conducted which is described by Eq.(5). These operations perform as non-interference accumulations which can filter out relatively large portions of noise as well according to the theory of probability and stochastic process.

**Conclusions and discussions.** In this letter, the fiber optical neural network and its related signal processing methods for the future fiber optical communication signal processing are proposed. Once being adopted, this neural network can further propel the deep integration between AI and optics. Through the demonstration of MFR, the fiber neural network can not only posses the ability of accurate predicting but also noise resisting.

Future research will focus on adopting this optical neural network to directly process the communication signals instead of the pre-extracted statistical information of the signals.

**Funding.** the Youth Fund of the National Natural Science Foundation (NSFC) of China.(62301275); Project of Key Laboratory of Radar Imaging and Mircowave Photonics (Nanjing University of Aeronautics and Astronautics), Ministry of Education. (NJ20230003); The Research Start-up Fund of Nanjing University of Posts and Telecommunications. (NY223032).

**Disclosures**. The authors declare no conflicts of interest.

**Data availability**. Data underlying the results presented in this paper are not publicly available but may be obtained from the authors upon reasonable request.

**References**
1. Shen Y, Harris N C., Skirlo S, *et al*. Nature Photonics **11**, 441 (2017).
2. Lin X, Rivenson Y, Yardimci N T, *et al*. Science **361**, 1004 (2018).
3. Feldmann J, Youngblood N, Wright C D, *et al*. Nature **569**, 208 (2019).
4. Sun Y, Dong M, Yu M, *et al*. Optics Letters **47**, 126 (2023).
5. Vandoorne K Dambre J Verstraeten D, *et al*. IEEE Transactions on Neural Networks **22**, 1649 (2011).
6. Duport F, Schneider B, Smerieri A, *et al*. Optics Express **20**, 22783 (2012).
7. Xu X, Tan M, Corcoran B, et al. Nature 5**89**, 44 (2021).
8. Wang D, Song Y, Li J, *et al*. Journal of Lightwave Technology **38**, 4730 (2020).
9. Yang H, Niu Z, Xiao S, *et al*. Journal of Lightwave Technology **39**, 1322 (2020).
10. Yubin Z, Zhenming Y, Kun X, *et al*. Optics Express **30**, 46626 (2022).
11. Yubin Z, Zhenming Y, Kun X, *et al*. Journal of Lightwave Technology **19,** 404 (2022)..
12. He X, Yan L, Jiang L, *et al*. Journal of Lightwave Technology **41**, 2301 (2022).
13. Yubin Z, Zhenming Y, Kun X, *et al*. Journal of Lightwave Technology **40,** 6347 (2022).
14. Meuris B , Qadeer S , Stinis P. DOI:10.48550/arXiv.2111.05307 (2021).
15. Karanov B, Chagnon M, Thouin F, *et al*. Journal of Lightwave Technology **36**, 4843 (2018).
16. Fang Y, Han H B, Bo W B, *et al*. Optics Letters **48**, 779 (2023)
17. Yubin Z, Minghua C, Sigang Y, *et al*. IEEE Journal of Selected Topics in Quantum Electronics **26**, 1 (2020).
18. Yubin Z, Minghua C, Sigang Y, *et al*. Science China Information Sciences, **64**, 2 (2021).

**References with titles include**

1. Shen Y, Harris N C., Skirlo S, *et al*. "Deep learning with coherent nanophotonic circuits," Nature Photonics **11**, 441 (2017).
2. Lin X, Rivenson Y, Yardimci N T, *et al*. "All-optical machine learning using diffractive deep neural networks," Science **361**, 1004 (2018).
3. Feldmann J, Youngblood N, Wright C D, *et al*. "All-optical spiking neurosynaptic networks with self-learning capabilities," Nature **569**, 208 (2019).
4. Sun Y, Dong M, Yu M, *et al*. "Modeling and simulation of all-optical diffractive neural network based on nonlinear optical materials," Optics Letters **47**, 126 (2023).
5. Vandoorne K Dambre J Verstraeten D, *et al*. "Parallel reservoir computing using optical amplifiers," IEEE Transactions on Neural Networks **22**, 1649 (2011).
6. Duport F, Schneider B, Smerieri A, *et al*. "All-optical reservoir computing," Optics Express **20**, 22783 (2012).
7. Xu X, Tan M, Corcoran B, et al. "11 TOPS photonic convolutional accelerator for optical neural networks," Nature 5**8**9, 44 (2021).
8. Wang D, Song Y, Li J, *et al.* "Data-driven optical fiber channel modeling: A deep learning approach," Journal of Lightwave Technology **38**, 4730 (2020).
9. Yang H, Niu Z, Xiao S, *et al*. "Fast and accurate optical fiber channel modeling using generative adversarial network," Journal of Lightwave Technology **39**, 1322 (2020).
10. Yubin Z, Zhenming Y, Kun X, *et al*. " Data-driven fiber model based on the deep neural network with multi-head attention mechanism,"Optics Express **30**, 46626 (2022).
11. Yubin Z, Zhenming Y, Kun X, *et al*. " Principle-driven fiber transmission model based on PINN neural network," Journal of Lightwave Technology **19,** 404 (2022)..
12. He X, Yan L, Jiang L, *et al*. "Fourier neural operator for accurate optical fiber modeling with low complexity,"Journal of Lightwave Technology **41**, 2301 (2022).
13. Yubin Z, Zhenming Y, Kun X, *et al*. "Multi-span long-haul fiber transmission model based on cascaded neural networks with multi-head attention mechanism," Journal of Lightwave Technology **40,** 6347 (2022).
14. Meuris B , Qadeer S , Stinis P. " Machine-learning custom-made basis functions for partial differential equations," DOI:10.48550/arXiv.2111.05307 (2021).
15. Karanov B, Chagnon M, Thouin F, *et al*. "End-to-end deep learning of optical fiber communications,"Journal of Lightwave Technology **36**, 4843 (2018).
16. Fang Y, Han H B, Bo W B, *et al. "*Deep neural network for modeling soliton dynamics in the mode-locked laser,*"* Optics Letters **48**, 779 (2023)
17. Yubin Z, Minghua C, Sigang Y, *et al*. "Electro-optical neural networks based on time-stretch method," IEEE Journal of Selected Topics in Quantum Electronics **26**, 1 (2020).
18. Yubin Z, Minghua C, Sigang Y, *et al*. "Optoelectronic convolutional neural networks based on time-stretch method," Science China Information Sciences, **64**, 2 (2021).